\documentclass[aps,prl,showpacs,twocolumn]{revtex4-1}
\usepackage{bm,color,amsmath,amssymb,mathrsfs,latexsym,graphicx,psfrag}

\newcommand{\bs}[1]{\boldsymbol{#1}}

\begin{document}
\title{Exotic $d$-wave superconductivity in strongly hole-doped
  $\text{K}_{x}\text{Ba}_{1-x}\text{Fe}_2\text{As}_2$ } 

\author{Ronny Thomale${}^1$} 
\author{Christian Platt${}^2$} 
\author{Werner Hanke${}^2$} 
\author{Jiangping Hu${}^3$}
\author{B.~Andrei Bernevig${}^1$} 

\affiliation{${}^1$Department of Physics, Princeton University, Princeton,
  NJ 08544}
\affiliation{${}^2$Institute for Theoretical
  Physics and Astrophysics, University of W\"urzburg, Am Hubland, D
  97074 W\"urzburg}
  \affiliation{${}^3$Department of Physics, Purdue University, West Lafayette, Indiana 47907, USA}


\begin{abstract}
  We investigate the superconducting phase in the
  $\text{K}_{x}\text{Ba}_{1-x}\text{Fe}_2\text{As}_2$ 122 compounds
  from moderate to strong hole-doping regimes. Using functional
  renormalization group, we show that while the system develops a
  nodeless anisotropic $s^{\pm}$ order parameter in the moderately
  doped regime, gapping out the electron pockets at strong hole doping
  drives the system into a nodal $(\cos k_x + \cos k_y)(\cos k_x -
  \cos k_y)$ $d$-wave superconducting state.  This
  is in accordance with recent experimental evidence from measurements
  on $\text{KFe}_2\text{As}_2$ which observe a nodal order parameter
  in the extreme doping regime. The magnetic instability is strongly
  suppressed.
\end{abstract}
\date{\today}


\maketitle

The most elementary questions in the field of iron-based
superconductors, such as the symmetry of the order parameter in the
superconducting (SC) state, are still under vivid debate.  The
complexities involve an intricate band structure, a diversity of
different material compounds which exhibit sometimes contradictory
behavior, and the proximity of various symmetry-broken phases. Due to
best single-crystal quality, the most studied pnictide compounds
belong to the 122 family such as $\text{BaFe}_2\text{As}_2$. Their
crystal structure is tetragonal I4/mmm, where the Fe and As atoms
arrange into layers; the intra-layer hybridization is dominant but,
unlike other pnictide compounds such as the 1111 family, the
inter-layer hybridization is also important.  Soon after their
discovery~\cite{kamihara-08jacs3296}, the 122 compounds have been
synthesized not only with Ba as a substituent between the FeAs layers,
but also with K, Cs, and Sr. The SC transition temperatures achieved
were up to $37$ K~\cite{sasmal-08prl107007}.

The current theoretical opinion on the SC order parameter has
converged on a nodeless $s^{\pm}$ order parameter that changes sign
between the electron ($e$) pockets and hole ($h$) pockets.  This order parameter comes out
of both the strong and the weak-coupling pictures of the iron-based
superconductors~\cite{seo-08prl206404,mazin-08prl057003,chubukov-08prb134512,stanev-08prb184509,maier-09prb224510},
and owes its origin to the pnictide Fermi surface (FS) topology of $h$ pockets at the $\Gamma$ and $e$ pockets at the $X$ $(\pi, 0)
/ (0,\pi)$ point of
the unfolded Brillouin zone. The dominant scattering contributions
originate from $h$ pocket scattering at $\Gamma$ to $e$ pockets at $X$,
yielding the $s^{\pm}$ SC order parameter for the doped case and the
collinear antiferromagnetic phase in the undoped case. Detailed
nesting properties of the pockets, the multi-orbital character of the
FS, and the presence or absence of a third $h$ pocket at $M$ $(\pi,\pi)$
in the unfolded Brillouin zone complicate this picture. For the 1111
compounds, it was shown that the absence of the $M$ $h$ pocket (whose
Fermi level can be significantly tuned by the pnictogen height through
replacing $\text{As}$ by $\text{P}$) can modify the SC order parameter
anisotropy from a nodeless to a nodal $s^{\pm}$
phase, which gives the correct material trend for $\text{As}$-$\text{P}$ substitution in other pnictide families~\cite{kuroki-09prb224511,thomaleasvsp,maier-09prb224510}.
With small exceptions, the anisotropic extended $s$-wave scenario (and
its extension to the nodal $s^{\pm}$) was consistent with experimental
findings for most of the pnictide
compounds~\cite{hashimoto-09prl207001,malone-09prb140501,tanatar-09cm0907,checkelsky-09cm0811,yamashita-09prb220509,hicks-09prl127003}.

It was realized at a very early stage that electron and hole doping
can have qualitatively different effects in the
pnictides~\cite{xu-08epl67015}. Hole doping should increase the
propensity to a nodeless ($s^{\pm}$) SC phase. The qualitative picture
applies to both the 122 and the 1111 compounds: as the Fermi level is
lowered, the $M$ $h$ pocket becomes more relevant and the $M
\leftrightarrow X$ scattering adds to the $(\pi,0)$/$(0,\pi)$
scattering from $\Gamma$ to $X$. As such the anisotropy-driving
scattering such as inter-electron pocket scattering becomes less
relevant and it yields a nodeless, less anisotropic and more stable $s^\pm$~\cite{thomaleasvsp,opti}. This picture is qualitatively confirmed by experiments. While
thermoelectric, transport and specific heat measurement have been
performed for $\text{K}_{x}\text{Ba}_{1-x}\text{Fe}_2\text{As}_2$ from
$x=0$ to the strongly hole-doped case
$x=1$~\cite{yan-10prb235107,ni-08prb014507,chen-08prb224512}, more
detailed studies have previously focused on the optimally doped case $x=0.4$
with $T_{\text{c}}=37$ K, where all measurements such as penetration
depth and thermal conductivity find indication for a moderately
anisotropic nodeless
gap~\cite{martin-09prb020501,checkelsky-09cm0811,luo-09prb140503,rotter-08prl107006}.
ARPES on doped $\text{BaFe}_2\text{As}_2$, likewise, finds a nodeless SC
gap~\cite{wray-08prb184508,ding-08epl47001,zhang2010d}.

The experimental findings for the SC phase in
$\text{KFe}_2\text{As}_2$ came as a surprise. Thermal
conductivity~\cite{dong-10prl087005,terashima-10prl259701},
penetration depth~\cite{hashimoto-10prb014526}, and
NMR~\cite{fukazawa-09jpsj083712,zhang-10prb012503} provide clear
indication for nodal SC. The critical temperature for
$\text{KFe}_2\text{As}_2$ is at $T_{\text{c}} \sim 3$ K, one order of
magnitude less than the optimally doped samples.  ARPES
measurements~\cite{sato-09prl047002} show that the $e$ pockets 
have nearly vanished, while the $h$
pockets at the folded $\Gamma$ point are large and have a linear
dimension close to $\pi/a$. In this Letter, we provide a detailed
picture of how the SC phase evolves under hole doping in
$\text{K}_{x}\text{Ba}_{1-x}\text{Fe}_2\text{As}_2$, and find that the
nodal phase observed for $x=1$ is of (extended) $d$-wave type. We use functional
renormalization group (FRG) to investigate how the SC form factor
evolves under doping from nodeless anisotropic $s^{\pm}$ in the
moderately hole-doped regime to $d$-wave in the strongly hole-doped
regime, where the $e$ pockets are assumed to be gapped out. The
$d$-wave SC minimizes the on-pocket hole interaction energy. We find
the critical divergence scale to be an order of magnitude lower than
for the optimally doped $s^{\pm}$ scenario, which is consistent with
experimental evidence.

\begin{figure}[t]
  \begin{minipage}[l]{0.99\linewidth}
    \includegraphics[width=\linewidth]{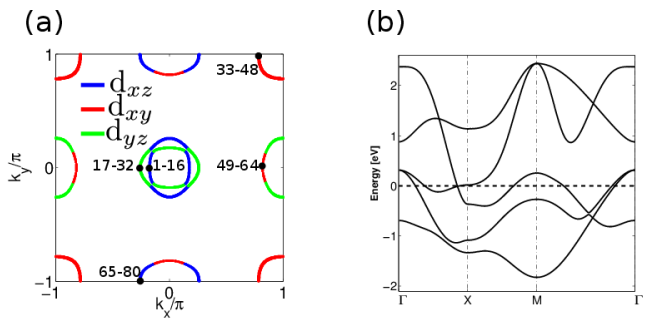}
  \end{minipage}
  \caption{(color online) (a) Schematic plot of the unfolded FS. (b)
    $k_z=0$ slice of the 122 band structure given in~\cite{graser-10prb214503}. The
  dominant orbital weights along the $e$ and $h$ pockets are
  highlighted. The patches along pockets are enumerated
  counterclockwise, starting at each pocket with the patch indicated by a
  dot. The number of patches with electron pockets is 80, and without it decreases to 48.}
\label{pic1}
\vspace{-0pt}
\end{figure}

\begin{figure*}[t!]
  \begin{minipage}[l]{0.70\linewidth}
    \includegraphics[width=\linewidth]{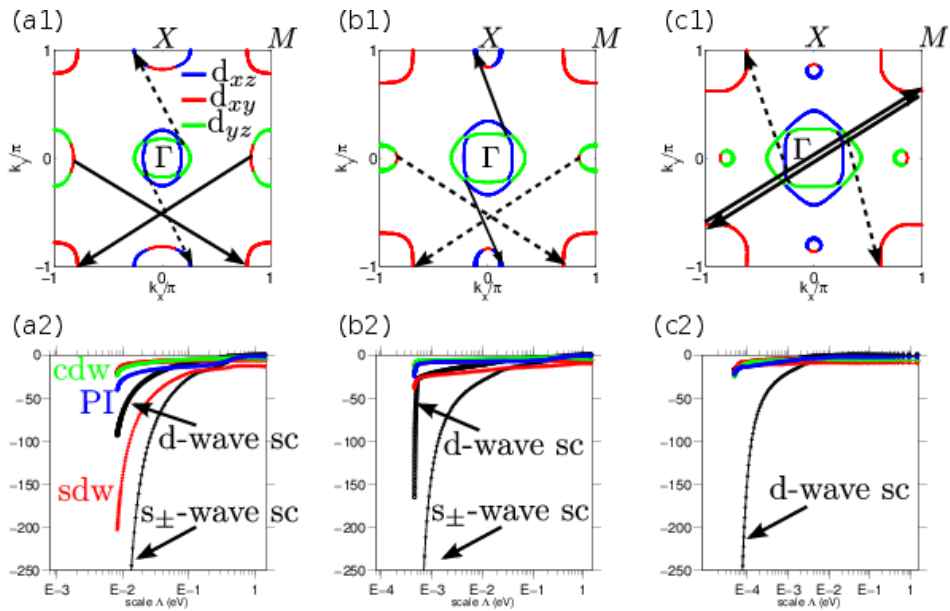}
  \end{minipage}
  \caption{(color online) Representative scenarios of the FS (unfolded
    BZ) and instability eigenvalue flows for chemical potential and
    electron concentration per iron $\mu=-.12, n_{\text{el}}=5.913$
    (a), $\mu=-.22, n_{\text{el}}=5.663$ (b), and $\mu=-.32,
    n_{\text{el}}=5.346$ in (c). The hole doping of our model
    calculation in (c),
    while exceeding the experimental setup $n_{\text{el}}=5.5$, best
    matches the FS profile from ARPES~\cite{sato-09prl047002}. The dominant and subdominant
    scatterings in the Cooper channel are highlighted in (a1)-(c1) by
    full and dashed arrows. The color contours along the FS label the
    dominant orbital weights [inset (a1)]. The leading eigenvalue flow
    of the ordering channel for different Fermi instabilities [charge
    density wave (CDW), Pomeranchuk (PI), spin density wave (SDW) and
    superconductivity (SC)] are plotted in (a2)-(c2) versus the IR
    cutoff FRG flow parameter $\Lambda$. For (a) and (b) we find
    $s^{\pm}$ as the leading Fermi instability. For (c) we observe a
    leading $d$-wave instability.}
 \label{pic2}
 \vspace{-0pt}
 \end{figure*}

We focus on studying
$\text{K}_{x}\text{Ba}_{1-x}\text{Fe}_2\text{As}_2$ starting at
the optimally doped case around $x=0.4$ and increasing the hole doping up to
$\text{KFe}_2\text{As}_2$. We use an effective 5-band tight-binding
model developed by Graser {\it et al.}~\cite{graser-10prb214503} to
describe the band structure of the 122-type iron-based
superconductors [see Fig.~\ref{pic1}]:
\begin{equation}
H_{0} =
\sum_{\bs{k},s}\sum_{a,b=1}^{5}c_{\bs{k}as}^{\dagger}K_{ab}(\bs{k})c_{\bs{k}as}^{\phantom{\dagger}}.
\label{bs}
\end{equation}
Here $c$'s denote electron annihilation operators, $a,b$ the five Fe
$d$-orbitals, $K$ the band matrix, and $s$ the spin index. As seen
in Fig.~\ref{pic1} and Fig.~\ref{pic2}, for moderate hole doping, the
conventional five pocket scenario with $e$ pockets at $X (\pi, 0)$
and $M (\pi,\pi)$ emerges. For larger hole doping, the $e$
pockets vanish and only small disconnected lobe features are found
around $X$ [Fig.~\ref{pic2}c]. The kinetic model reduces to the
effectively three $h$ pocket scenario shown in Fig.~\ref{pic2}c.  Other
details of the 122 band structure are currently still under debate, with
unresolved questions about the FS topology at the $Z$ point
in the three dimensional Brillouin zone and the importance of
integrating over the full range along $k_z$.
 However, as many of these details mostly affect
the $e$ pocket anisotropies, they are irrelevant for our proposed SC mechanism: as we always
consider a rather largely hole-doped regime, the $e$ pockets can
be assumed relatively small - even disappearing in the most
interesting case, i. e. that of full hole doping.  We, therefore, 
particularize to the $k_z=0$ cut of~\eqref{bs} in the following, and
also omit the lobe features at large hole doping within the RG calculations.  To
test our assumption of the irrelevance of the $k_z$ dispersion to our
results, we have made several other cuts at different $k_z$ and
confirmed that our results do not change qualitatively. 
We cannot ultimately exclude that the lobes may
influence the system due to the fact that our Brillouin zone patching
scheme is not fully adequate for such small Fermi surface
features. Still, within our formalism, we find that the lobes are negligible in the RG flow.


A schematic picture of the FS topology is given in
Fig.~\ref{pic1}a. The $h$ pockets at $\Gamma$ mainly
have $d_{xz}$ and $d_{yz}$ orbital content while the $h$ pocket at $M$ consists of
$d_{xy}$ orbital weight. When present, the $e$ pockets consist of $d_{xz}$ and $d_{yz}$
orbital weight. Exceeding a certain size, the front tip 
along $\Gamma - X$ also has an important $d_{xy}$ weight on the $e$ pockets. 
We use the conventional onsite orbital model for the interactions, i. e.
\begin{eqnarray}
&&H_{\text{int}}=\sum_i \left[ U_1 \sum_{a} n_{i,a\uparrow}n_{i,a\downarrow} + U_2\sum_{a<b,s,s'} n_{i,as}n_{i,bs'} \right.\nonumber \\
&& \hspace{-15pt}\left.+\sum_{a<b}(J_\text{H}\sum_{s,s'} c_{ias}^{\dagger}c_{ibs'}^{\dagger}c_{ias'}^{\phantom{\dagger}}c_{ibs}^{\phantom{\dagger}}  +J_{\text{pair}} c_{ia\uparrow}^{\dagger}c_{ia\downarrow}^{\dagger}c_{ib\downarrow}^{\phantom{\dagger}}c_{ib\uparrow}^{\phantom{\dagger}}) \right]\hspace{-4pt},
\end{eqnarray}
where $n_{i,as}$ denote density operators at site $i$ of spin $s$ in
orbital $a$. We consider intra- and inter-orbital interactions $U_1$
and $U_2$ as well as Hund's coupling $J_{\text{H}}$ and pair hopping
$J_{\text{pair}}$. 
We choose the values of the interaction parameters close to the ones obtained by constrained
RPA ab initio calculations~\cite{imada}: $U_1 > U_2 > J_\text{H} \sim
J_{\text{pair}}$, and set $U_1 = 3.0 eV, U_2 = 2.0 eV,
J_{\text{H}}=J_{\text{pair}}=0.6 eV$. While there are variations of
these parameters for different classes of pnictides, the values of the
parameters are all in the same range, and we have confirmed 
that variations of 20-30 \% of the interaction parameters do not
change the picture qualitatively. As a tendency, a comparably large
absolute value of $U_1$ needs to be kept to trigger the SC instability,
where increasing $U_2$ also helps to increase the critical cutoff
scale and, thus, $T_\text{c}$.


Using multi-band FRG~\cite{wang-09prl047005,thomale-09prb180505,thomaleasvsp,platt-09njp055058},
we study the evolution of the renormalized interaction described by the 4-point
function (4PF) under integrating out high energy fermionic
modes:
$V_{\Lambda}(\bs{k}_1;\bs{k}_2;\bs{k}_3;\bs{k}_4)c_{\bs{k}_4s}^{\dagger}c_{\bs{k}_3\bar{s}}^{\dagger}c_{\bs{k}_2s}^{\phantom{\dagger}}c_{\bs{k}_1\bar{s}}^{\phantom{\dagger}},$
where the flow parameter is the IR cutoff $\Lambda$ approaching the
FS. $\bs{k}_{1}$ to $\bs{k}_{4}$ are the incoming and
outgoing momenta.  
The starting conditions
are given by  the bare
initial interactions for the 4PF with the bandwidth serving as an UV cutoff. The diverging channels of the 4PF
under the flow to the FS signal the nature of the
instability, and the corresponding $\Lambda_{\text{c}}$ serves as an upper
bound for the transition temperature $T_{\text{c}}$. The Cooper
channel of the 4PF provides the different SC form factors - the dominant order parameter having the largest eigenvalue ~\cite{wang-09prl047005,thomale-09prb180505,thomaleasvsp,platt-09njp055058}. In Fig.~\ref{pic2},
the leading eigenvalues for different FS instabilities are plotted against $\Lambda$ for different fillings between moderately hole-doped from the left to strongly
hole-doped to the right. We find that for all scenarios the leading instability is in the Cooper
channel.

For the moderately doped case, the $e$ pockets are of similar size as
the $h$ pockets. Fig.~\ref{pic2} (a1) shows the FS structure as
well as the dominant (full line) and subdominant scattering (dashed
arrow) processes in the Cooper channel. The two major components are
given by $\Gamma \leftrightarrow X$ as well as $M
\leftrightarrow X$ scatterings. They are particularly important for the
front tips of the $e$ pockets since these parts can scatter to $M$ via
dominant $U_1$ interaction due to identical orbital content. The SDW
fluctuations are strong, signaling the proximity
to the leading magnetic instability scenario of the undoped model
[Fig.~\ref{pic2} (a2)].

\begin{figure}[t]
  \begin{minipage}[l]{0.99\linewidth}
    \includegraphics[width=\linewidth]{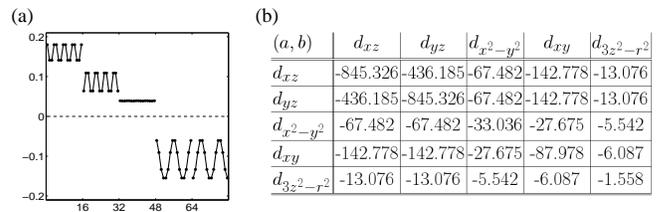}
  \end{minipage}
  \caption{(color online) (a) Form factor of the leading SC
    instability of scenario Fig.~\ref{pic2}b, plotted versus the
    patching index of the Fermi pockets according to
    Fig.~\ref{pic1}a. (b) Eigenvalues of the orbital decomposition of
    the SC form factor in~(a). 
The ratio of the values label the
    relative importance of the orbital scattering channel $V(a,a
    \rightarrow b,b) c_{a \uparrow}c_{a \downarrow}c_{b\downarrow}^{\dagger}c_{b\uparrow}^{\dagger}$.}
\label{pic3}
\vspace{-0pt}
\end{figure}

For the intermediate regime, between moderate and strong hole doping,
the $e$ pockets are already very small [Fig.~\ref{pic2}
(b1)]. The nesting to the $h$ pocket is absent, and the SDW
fluctuations are strongly reduced. In addition, the SDW fluctuations
become less concentrated in the $(\pi,0)/(0,\pi)$ or $(\pi,\pi)$
channel, and spread into various incommensurate sectors~\cite{Lee}. The $d_{xy}$ orbital weight on the
$e$ pocket is
reduced and the $M \leftrightarrow X$ scattering becomes
subdominant. The main Cooper channel scattering is along
$\Gamma \leftrightarrow X$. As a consequence, $s \pm$ is still the
leading instability, where the form factor and its decomposition into
orbital scattering contributions are shown in Fig.~\ref{pic3}: the
largest gap is found for the inner $h$ pocket at $\Gamma$, followed by
the outer $h$ pocket and the $h$ pocket at $M$, where the $e$ pockets show
anisotropic gaps. The orbital decomposition confirms the previous
discussion of the the dominant scattering contribution, in that the largest
weight resides at intra and inter-orbital scattering of the $d_{xz}$
and $d_{yz}$ orbital.  However, we already observe that, due to the
lack of SDW fluctuations supporting the SC, the critical divergence
scale is decreased [Fig.~\ref{pic2} (a2)-(c2)]. In particular, while still subdominant, we can
already see the $d$-wave evolving as the second-highest instability
eigenvalue in the Cooper channel. When $e$
pockets are still present, the form factor
(not shown here)  closely resembles the extended $d$-wave type
involving $h$ pockets and
$e$ pockets~\cite{thomale-09prb180505}. 

\begin{figure}[t]
 \begin{minipage}[l]{0.99\linewidth}
   \includegraphics[width=\linewidth]{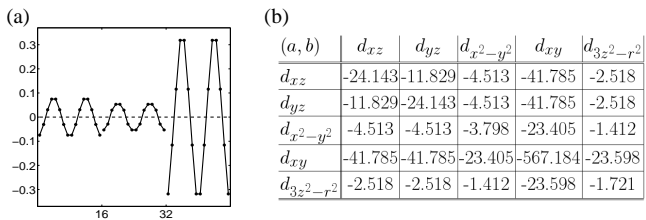}
  \end{minipage}
 \caption{(color online) (a) Form factor of the leading SC
    instability of the strongly hole-doped scenario Fig.~\ref{pic2}c, plotted versus the
    patching index of the FS according to
    Fig.~\ref{pic1}a. (b) Eigenvalues of the orbital decomposition of
    the SC form factor in~(a). We find a extended $d$-wave form factor. The nodal
  points are given along the main diagonal of the Brillouin zone, i.e.
the form factor of the inner $h$ pocket, for example, crosses zero
between the patches $(2,3)$, $(6,7)$, $(10,11)$, and $(14,15)$.}
\label{pic4}
\vspace{-0pt}
\end{figure}

At strong hole doping, the $e$ pockets are absent, and the $h$ pockets
are very large. The flow in Fig.~\ref{pic2}c shows no instability up
to small cutoff scales of $\Lambda$ where we find a leading
instability in the Cooper channel. Its form factor and orbital
scattering decomposition is shown in Fig.~\ref{pic4}. We observe an extended
$d$-wave instability on the three $h$ pockets, with nodes located
along the main diagonals in the Brillouin zones [as seen by comparing
the patch numbers of the $0$ crossings in Fig.~\ref{pic4}a and the
patching enumeration defined in Fig.~\ref{pic1}].  A harmonic analysis
of the
order parameter yields a large contribution of $\cos(2k_x)-
\cos(2k_y)$ type and a subdominant $\cos (k_x) - \cos (k_y)$
component, i.e. the form factor is most accurately characterized as $(\cos k_x + \cos k_y)(\cos k_x -
  \cos k_y)$. The dominant
scattering is intra-pocket scattering on the large $M$  $h$ pocket, followed
by inter-orbital $d_{xy}$ to $d_{xz,yz}$ scattering between $M
\leftrightarrow \Gamma$. While the magnetic fluctuations are generally
weak in this regime, the dominant contribution is now given by
$(\pi,\pi)$ SDW fluctuations as opposed to $(\pi,0) / (\pi,0)$ for
smaller hole doping. For strong
hole doping, the $h$ pocket at $M$ is large enough to induce
higher harmonic $d$-wave SC through intra-pocket scattering between the $d_{xy}$
orbitals as confirmed by the large value of $d_{xy}-d_{xy}$ pairing [Fig.~\ref{pic4}b]. Via scattering to the other pockets, the SC
instability is likewise induced there, however, with smaller amplitude
than for the $M$ pocket [Fig.~\ref{pic4}a]. As opposed to
  conventional first harmonic $d$-wave, there is no sign change
  between the extended $d$-wave form factor on the $M$ pocket and the
  $\Gamma$ pocket according to $\cos(2k_x)-
\cos(2k_y)$ [Fig.~\ref{pic4}]. This picture of a
$k$-space proximity effect from the $M$ pocket to the $\Gamma$ pockets is substantiated by our checks with
calculations involving the $M$ pocket only, where we see a similar
evolution of an SC instability (the divergence is lower, as the inter-orbital scatterings in the 3-pocket scenario
help to renormalize the repulsive Coulomb interactions).
This matches the orbital
decomposition of the SC form factor in Fig.~\ref{pic4}b, showing
dominant intra-orbital scattering of the $d_{xy}$ orbital. 


As apparent from
the ARPES data, the nodal character of the SC phase in
$\text{KFe}_2\text{As}_2$ cannot originate from possible nodes on the
$e$ pockets (which are gapped out at these doping levels) 
but must be due to nodes on the $h$ pockets. It is then clear that the
order parameter cannot be $s^\pm$ as it does not tend to allow for an anisotropy that would drive the $h$ pockets nodal. The $d$-wave
instability which we find for the strongly hole-doped regime provides
an explanation for the general experimental evidence, while the
detailed gap structure certainly deserves further investigation~\cite{Kawano-Furukawa}. Electron-phonon coupling
may change the picture slightly quantitatively, but not qualitatively,
as the nodal features tentatively linked to the $d$-wave symmetry are
unambiguously observed in experiment. Pnictogen height variations as a
function of doping may change the precise value of $T_c$, and
would be important to be studied in general from first principles. Finally, it would be
interesting to further analyze how the system evolves from the
$s$-wave SC phase to the $d$-wave SC phase as a function of doping.

\begin{acknowledgments}
  We thank A.~Chubukov, H.~Ding, S.~Graser, D.~van~Harlingen,
  P.~Hirschfeld, D.~Scalapino, and Y.-L.~Wu for discussions. RT and BAB thank the IOP
  Beijing for hospitality, where some of the work has been done. We
  thank all participants of the KITP workshop 'Iron-based
  superconductors' for discussions. We
  acknowledge computational 
resources by Reed College. RT is
  supported by the Humboldt Foundation.  RT and CP are supported by
  DFG-SPP 1458/1, CP by DFG-FOR 538. BAB was supported by Sloan Foundation, NSF DMR- 095242, NSF China 11050110420, and MRSEC grant at Princeton University, NSF DMR-0819860.
\end{acknowledgments}

\end{document}